\documentclass[12pt]{article}%
\usepackage{t1enc}
\usepackage{amsmath}
\usepackage{graphicx}%
\usepackage{amsfonts}%
\usepackage{amssymb}
\newtheorem{theorem}{Theorem}

\newtheorem{corollary}[theorem]{Corollary}

\newtheorem{lemma}[theorem]{Lemma}

\newenvironment{proof}[1][Proof]{\noindent\textbf{#1.} }{\ \rule{0.5em}{0.5em}}

\begin{document}

\author{Garry Ludwig\thanks{Department of Mathematical Sciences, University of
Alberta, Edmonton, Alberta, Canada T6G 1S6}
\and S. Brian Edgar\thanks{Department of Mathematics, University of Link\"{o}ping,
Link\"{o}ping, Sweden S-581 83.}}
\title{(Conformal) Killing vectors in the Newman-Penrose formalism}
\date{June 7, 2001}
\maketitle

\begin{abstract}
Finding (conformal) Killing vectors of a given metric can be a difficult task.
This paper presents an efficient technique for finding Killing, homothetic, or
even proper conformal Killing vectors in the Newman-Penrose (NP) formalism.
Leaning on, and extending, results previously derived in the GHP formalism we
show that the (conformal) Killing equations can be replaced by a set of
equations involving the commutators of the Lie derivative with the four NP
differential operators applied to the four coordinates.

It is crucial that these operators refer to a preferred tetrad relative to the
(conformal) Killing vectors, a notion to be defined. The equations can then be
readily solved for the Lie derivative of the coordinates, i.e. for the
components of the (conformal) Killing vectors. Some of these equations become
trivial if some coordinates are chosen intrinsically (where possible), i.e. if
they are somehow tied to the Riemann tensor and its covariant derivatives.

If part of the tetrad, i.e.\thinspace part of null directions and gauge, can
be defined intrinsically then that part is generally preferred relative to any
Killing vector. This is also true relative to a homothetic vector or a proper
conformal Killing vector provided we make a further restriction on that
intrinsic part of the tetrad. If because of null isotropy or gauge isotropy,
where part of the tetrad cannot even in principle be defined intrinsically,
the tetrad is defined only up to (usually) one null rotation parameter and/or
a gauge factor, then the NP-Lie equations become slightly more involved and
must be solved for the Lie derivative of the null rotation parameter and/or of
the gauge factor as well. However, the general method remains the same and is
still much more efficient than conventional methods.

Several explicit examples are given to illustrate the method.

\end{abstract}

\section{Introduction}

Within tetrad formalisms there is a standard procedure for finding (conformal)
Killing vectors ((C)KVs), i.e. Killing vectors (KVs), a homothetic vector
(HV), and/or proper conformal Killing vectors (CKVs) for a given spacetime or
for specifying their existence while solving Einstein's equations for the
spacetime. This procedure, which includes integration of the tetrad version of
the (conformal) Killing equations, is, however, long and complicated as well
as inefficient due to several redundancies in the equations.

Recently\cite{EL1,EL2}, we have developed a new method for obtaining an HV
and/or KVs for spacetime solutions found in the Geroch-Held-Penrose\cite{GHP}
(GHP) formalism. We proved that the (conformal) Killing equations can be
replaced as key equations by a set of sixteen commutator equations involving
the four GHP differential operators and a new generalized Lie derivative
operator (called the GHP Lie derivative). This procedure can, with some care,
be extended to CKVs as well.

In this paper we adapt our method to the better-known, if more cumbersome,
Newman-Penrose\cite{NP} (NP) formalism. We shall replace the (conformal)
Killing equations by a set of sixteen commutator equations involving the four
NP derivative operators and the \emph{ordinary} Lie derivative. In principle,
to find HV, KVs and CKVs we only need to construct an appropriate null tetrad
for a given metric. However, the calculations can often be shortened
considerably if we know something about the geometric nature of the tetrad and
the coordinates used. We usually have such knowledge if we are in the process
of finding a spacetime solution possessing certain symmetries. But even for a
given metric we may calculate, for example, the Weyl tensor, and use one or
both of its principal null directions as the tetrad's null directions. As we
shall see, such knowledge, though not necessary in principle, usually reduces
significantly the complexity of the new equations.

The main aim of this paper is to address the problem of finding (C)KVs, i.e.
an HV, KVs, and CKVs, using the NP formalism, and to illustrate our method
with explicit examples.

One of the main differences between the NP and GHP formalisms is that the
former uses specific null tetrads whereas the latter deals only with classes
of null tetrads (called GHP tetrads here), where only the directions of the
two real null vectors are fixed but not the gauge. The gauge freedom consists
of the ability to make spin- and boost transformations.

In a previous paper \cite{EL2} we defined the notion of an intrinsic GHP
tetrad, as well as that of a preferred GHP tetrad relative to a vector, in
particular relative to a (C)KV. We argued that an intrinsic GHP tetrad was
preferred relative to any KV. (For an HV or a CKV the situation is a bit more
complicated). We proved the equivalence of the usual (conformal) Killing
equations with the GHP-Lie commutator equations, equations involving the
commutators of a newly defined generalized Lie derivative, called the GHP Lie
derivative, with the four GHP differential operators corresponding to a
preferred GHP tetrad, applied to the four coordinates.

In this paper, we define the notion of a preferred gauge relative to a (C)KV
as one in which the GHP Lie derivative and the ordinary Lie derivative
coincide. As we shall see, such a gauge always exists. The (conformal) Killing
equations can then be replaced by commutator equations involving commutators
of the ordinary Lie derivative and the NP differential operators applied to
the four coordinates, provided that the tetrad is preferred relative to the
(C)KV involved.

The question is then how to obtain such a preferred tetrad, especially when
the metric possesses more than one (C)KV and we want to find all of them at
once. One way is to leave enough arbitrariness in the tetrad so that the
commutator equations can choose a tetrad appropriate to each (C)KV. However,
that may complicate the equations somewhat$^{\text{F}}$\footnote{This is not
necessarily so. There was no such complication in the example of the pure
radiation metric of Ref.\cite{EL2}.}. For KV it is usually possible (and most
efficient) to find an intrinsic tetrad, i.e. one that can be defined in terms
of the Riemann tensor and its covariant derivatives. Such a tetrad is then
generally preferred relative to any KV. In case of null isotropy or gauge
isotropy we may have to be satisfied with a partially intrinsic tetrad. If we
want to include an HV or a CKV in our search, it is again best to ensure
(where possible) that the tetrad is intrinsic. However, in addition, the
conditions defining the null directions and the gauge must be properly
conformally weighted. For an HV, it suffices if there is a proper conformal
weight relative to a constant conformal transformation.

In section 2, we review the GHP approach to KVs and HV, as developed in
Ref.\cite{EL2}, and extend the results to CKVs. In section 3 we adapt these
results to the NP formalism. Several important lemmas will lead to the key
theorem involving the NP-Lie commutator equations. In section 4, we discuss
and illustrate our method further, with examples of metrics quoted in
canonical tetrads. Some appropriate remarks in section 5 will conclude the paper.

The NP version of our method is deduced in section 3 from our earlier GHP
version\cite{EL2}. Although we could have obtained these NP results
independently and entirely in the NP formalism it was much quicker to modify
the GHP results. However, the basic theorems at the end of Section 3, and the
theorems we deduce and use in Section 4 can be understood without any detailed
knowledge of the GHP formalism.

One of our motivations for this work is to be able to apply our results to the
invariant/Karlhede classification (IC) program\cite{Karl} for spacetimes. In
that program there exist algorithms to determine respectively the dimension of
the isometry group\cite{MK} of the spacetime and to determine whether the
spacetime admits a homothety\cite{KS}; these algorithms do not give the forms
of the symmetry groups, nor the explicit symmetry vectors. Our intention is to
fill this gap in a subsequent paper, by developing our method to find an HV
and KVs explicitly and efficiently from the information supplied by the IC of
a metric. We shall therefore use notation and conventions in this paper which
are consistent with the IC program.

There are a number of general results\cite{Hall1,Hall2} which give
restrictions on a CKV; in particular, it is known that the only vacuum
spacetimes admitting a CKV are the plane waves. Such theorems can be used to
simplify our calculations in specific applications.

Today there are a number of computer packages which calculate symmetries in
given spacetimes. It might be thought that these would replace existing
methods and, indeed, any calculations done by hand. However, it should be
recognized that these packages are still relatively limited. For instance,
recently, one package was unable to obtain results directly, results
pertaining to some symmetries of the Kerr metric\cite{Prince}, even though, as
pointed out in Ref.\cite{Hall1}, these particular symmetries were known not to
exist as a consequence of some general theorems. On the other hand, in section
4 of the present paper, our methods establish directly and quickly the
non-existence of an HV and CKVs for the Kerr-Newman metric. It is our
intention eventually to apply our methods not only to the classification and
calculation of symmetries of existing metrics but to use these methods as an
efficient means of imposing symmetries in a coordinate-invariant manner in the
search for new solutions.

\section{Preferred GHP tetrads and the GHP-Lie commutator equations}

Kolassis and Ludwig\cite{KL} introduced a generalized Lie derivative
$L_{_{\xi}}$ which is defined like the ordinary Lie derivative $\pounds
_{_{\xi}}$ except that ordinary derivatives are replaced by covariant
derivatives. Equivalently,%

\begin{equation}
L_{_{\xi}}=\pounds_{_{\xi}}-\xi^{^{\mu}}\left(  p\zeta_{_{\mu}}+q\overline
{\zeta}_{_{\mu}}\right)  , \label{A1}%
\end{equation}
where ($p,q$) are the GHP weights of the quantity operated upon, and
\begin{equation}
\zeta_{_{\mu}}=\gamma l_{_{\mu}}+\varepsilon n_{_{\mu}}-\alpha m_{_{\mu}%
}-\beta\overline{m}_{_{\mu}}, \label{A2}%
\end{equation}
where $\mathbf{l}$, $\mathbf{n}$, $\mathbf{m}$, $\overline{\mathbf{m}}$ are
null tetrad vectors (normalized by $l_{_{\mu}}n^{^{\mu}}=1,$ $m_{_{\mu}%
}\overline{m}^{^{\mu}}=-1$, other inner products zero), and $\alpha$, $\beta$,
$\gamma$, $\varepsilon$ are the ''bad'' Newman-Penrose spin-coefficients that
disappear from the GHP formalism.

A more useful operator turned out to be a further generalization, namely
\begin{equation}
\text{\L }_{_{\xi}}=L_{_{\xi}}-\frac{p}{4}\left(  \mathcal{P}-\mathcal{P}%
^{\prime}+\mathcal{P}^{\ast}-\mathcal{P}^{\prime\ast}\right)  -\frac{q}%
{4}\left(  \mathcal{P}-\mathcal{P}^{\prime}+\mathcal{P}^{\prime\ast
}-\mathcal{P}^{\ast}\right)  ,\label{A3}%
\end{equation}
where
\begin{equation}
\mathcal{P}=n_{_{\mu}}L_{_{\xi}}l^{^{\mu}},\label{A4}%
\end{equation}
and $\mathcal{P}^{\prime},\mathcal{P}^{\ast},$ and $\mathcal{P}^{\prime\ast}$
are its three companions\cite{KL}, i.e. its images under the prime, the star,
and the prime-star operations. An equivalent definition of this GHP Lie
derivative is$^{\text{F}}$\footnote{\L $_{_{\xi}}\,$as defined here reduces to
that of Eq.(4) of Ref.\cite{EL2} only in the case of a KV. This should have
been made more explicit there.}
\begin{align}
\text{\L }_{_{\xi}} &  =\pounds _{_{\xi}}+\left(  \frac{p+q}{4}\right)
\left(  l_{_{\mu}}\pounds _{_{\xi}}n^{^{\mu}}-n_{_{\mu}}\pounds _{_{\xi}%
}l^{^{\mu}}\right)  \nonumber\\
&  +\left(  \frac{p-q}{4}\right)  \left(  \overline{m}_{_{\mu}}\pounds _{_{\xi
}}m^{^{\mu}}-m_{_{\mu}}\pounds _{_{\xi}}\overline{m}^{^{\mu}}\right)
.\label{A5}%
\end{align}

Let us define the remaining tetrad components $\mathcal{Q}$ and $\mathcal{R}$
of \L $_{_{\xi}}l^{^{\mu}}$ by
\begin{equation}
\text{\L }_{_{\xi}}l^{^{\mu}}=\frac{1}{2}\left(  \mathcal{P}+\mathcal{P}%
^{\prime}\right)  l^{^{\mu}}+\mathcal{R}n^{^{\mu}}-\mathcal{Q}m^{^{\mu}%
}-\overline{\mathcal{Q}}\overline{m}^{^{\mu}},\label{A6}%
\end{equation}
and similarly for their companions under the the prime, star and star-prime
operations. A GHP tetrad is said to be preferred$^{\text{F}}$\footnote{The
motivation for this definition is not entirely obvious here. For an
explanation the reader is referred to Ref.\cite{EL2}.} relative to the vector
$\xi$ provided that
\begin{equation}
\mathcal{Q}^{\prime}+\mathcal{Q}^{\ast}=0=\mathcal{Q}+\mathcal{Q}^{\prime\ast
}.\label{A7}%
\end{equation}
Such a GHP tetrad always exists but is not unique; it allows for null
rotations with parameters $y$ and $z$ satisfying
\begin{equation}
\text{\L }_{_{\xi}}y=0=\text{\L }_{_{\xi}}z.\label{A8}%
\end{equation}

If, in addition, the (conformal) Killing equations hold, namely
\begin{align}
\mathcal{P}^{\prime}  &  =-\mathcal{P}-\varphi\text{, \hspace{0.6cm}%
}\mathcal{P}^{\prime*}=-\mathcal{P}^{*}-\varphi\nonumber\\
\mathcal{Q}  &  =\mathcal{Q}^{\prime*}\text{, \hspace{0.6cm}}\mathcal{Q}%
^{\prime}=\mathcal{Q}^{*}\nonumber\\
\mathcal{R}  &  =\mathcal{R}^{\prime}=\mathcal{R}^{*}=\mathcal{R}^{\prime*}=0
\label{A9}%
\end{align}
then
\begin{equation}
\text{\L}_{_{\xi}}l^{^{\mu}}=-\frac{1}{2}\varphi l^{^{\mu}}, \label{A10}%
\end{equation}
and similarly for its companions. A tetrad satisfying Eqs.(\ref{A10})
(including the companions) is said to be GHP Lie recurrent (\emph{GLR}).

Note again that the (C)KV is an HV if the function $\varphi$ is a constant and
a KV if $\varphi$ is zero; otherwise it is a CKV.

Although in the above we defined the notion of preferred null directions, i.e.
of a preferred GHP tetrad, for an arbitrary vector $\xi$, its main usefulness
is relative to a (C)KV. When the null directions are preferred relative to a
(C)KV, the generalized Lie derivative of such a GHP tetrad then takes its
simplest form, given by Eq.(\ref{A10}) and companion equations. In particular,
relative to a KV, a preferred GHP tetrad gets annihilated by the GHP Lie
derivative \L. We say it is GHP Lie derived.

More generally, for a (C)KV $\mathbf{\xi}$ and corresponding conformal factor
$\varphi$, if a quantity $\eta$ satisfies
\begin{equation}
\left(  \text{\L}_{_{\xi}}-w\frac{\varphi}{2}\right)  \eta=0, \label{A10a}%
\end{equation}
for some number $w$, we say that it is \emph{GLR} (relative to $\xi$ and with
associated factor $\varphi$) and that it has $w$-weight $w.$ It is not hard to
show that if $\eta$ is properly conformally weighted then the $w$-weight
coincides with the conformal weight. For the contravariant tetrad vectors and
for spin-coefficients well-behaved in the conformally extended GHP
formalism\cite{L1}, $w=-1$; for Weyl tensor tetrad components, $w=-2$. If,
again with $\mathbf{\xi}$ a (C)KV and $\varphi$ the corresponding conformal
factor, a quantity $\eta$ satisfies%

\begin{equation}
\left(  \pounds_{_{\xi}}-w\frac{\varphi}{2}\right)  \eta=0 \label{A10b}%
\end{equation}
we say that $\eta$ is (ordinary) \emph{Lie recurrent} (relative to the vector
$\mathbf{\xi}$ and with $w$-weight $w$).\medskip

\medskip\negthinspace\negthinspace\textbf{Definition 1.}

\medskip The \emph{GHP scalars with respect to a GHP tetrad} are the
well-behaved GHP spin-coefficients, all the Riemann tensor tetrad components,
together with all properly GHP weighted combinations of these and their GHP derivatives.\medskip

In Ref.\cite{EL2} we proved the following lemmas.

\begin{lemma}
If the GHP tetrad is GLR relative to a vector $\xi$ and with associated factor
$\varphi$ then $\xi$ is a (C)KV with $\varphi$ the conformal factor, and the
GHP tetrad is necessarily preferred relative to $\xi$.
\end{lemma}

\begin{lemma}
If $\xi$ is a (C)KV with conformal factor $\varphi$ then the GHP tetrad is GLR
relative to $\xi$ and with associated factor $\varphi$ iff it is preferred
relative to $\xi$.
\end{lemma}

\begin{lemma}
If the GHP tetrad is preferred relative to a vector $\xi$ then it is GLR
relative to $\xi$ with associated factor $\varphi$ iff $\xi$ is a (C)KV with
conformal factor $\varphi$.
\end{lemma}

\begin{lemma}
Given an HV or a KV $\xi$, all properly GHP weighted combinations of the GHP
tetrad, the GHP scalars, and the tetrad components of $\xi$ itself are GLR
relative to $\xi$ provided that the tetrad is preferred relative to $\xi$.
\end{lemma}

\begin{lemma}
If the GHP tetrad is GLR relative to a vector $\xi$ then there exists a
particular gauge in which the tetrad is (ordinary) Lie recurrent relative to
$\xi.$
\end{lemma}

\begin{lemma}
Relative to an HV or a KV, all GHP scalars of type (0,0) are (ordinary) Lie
recurrent in an arbitrary gauge of a preferred GHP tetrad.
\end{lemma}

\smallskip Note that the last two lemmas are already in a form suitable for
the NP formalism.

We now wish to generalize Lemma 4 to include \emph{proper} conformal Killing
vectors, i.e. CKVs. We know from Eqs.(27)-(35) and their companions in
Ref.\cite{EL2} that most GHP scalars which are \emph{GLR} relative to an HV or
a KV in a tetrad preferred with respect to that HV or KV, are also \emph{GLR}
relative to a CKV in a tetrad preferred with respect to that CKV, e.g.,
$($\L $_{_{\xi}}+\frac{1}{2}\varphi)\kappa=0.$ However, there are some GHP
scalars which are not \emph{GLR }relative to a CKV. For example,
$($\L $_{_{\xi}}+\frac{1}{2}\varphi)\overline{\pi}=\frac{1}{2}$\dh $\varphi$
and $($\L $_{_{\xi}}+\frac{1}{2}\varphi)\tau=-\frac{1}{2}$\dh $\varphi.$ It
follows easily that $(\overline{\pi}+\tau)$ is \emph{GLR} since $($%
\L $_{_{\xi}}+\frac{1}{2}\varphi)(\overline{\pi}+\tau)=0.$ Similarly, we can
show that \emph{all} properly conformally weighted GHP scalars are \emph{GLR}.

So we can deduce the general result,

\begin{lemma}
Given a (C)KV $\xi,$ all properly conformally weighted and properly GHP
weighted combinations of the GHP tetrad, the GHP scalars and the tetrad
components of $\xi$ itself are \emph{GLR} relative to $\xi,$ provided that the
GHP tetrad is preferred relative to $\xi$.
\end{lemma}

\smallskip

To be explicit\cite{L1}, the properly conformally weighted spin coefficients
(all with conformal weight -1) are, $\kappa,\sigma,\lambda,\nu,\rho
-\overline{\rho},\mu-\overline{\mu},\tau+\overline{\pi}$. The tetrad
components of the Weyl tensor are, of course, properly conformally weighted
(all with conformal weight -2). Although the Ricci tensor tetrad components by
themselves are not, the following combinations are properly conformally
weighted (with conformal weight -2),

$\Phi_{_{00}}+\frac{1}{2}\left[  -\text{\th}\left(  \rho+\overline{\rho
}\right)  +\frac{1}{2}\left(  \rho+\overline{\rho}\right)  ^{2}+\kappa\left(
\pi-\overline{\tau}\right)  +\overline{\kappa}(\overline{\pi}-\tau)\right]  $
(and companions),

$\Phi_{_{01}}+\frac{1}{2}\left[  \text{\th}(\overline{\pi}-\tau)+\frac{1}%
{2}(\overline{\pi}+\tau)\left(  \rho+\overline{\rho}\right)  +\kappa\left(
\mu+\overline{\mu}\right)  \right]  $ (and companions),

$\Phi_{_{11}}+\frac{1}{4}\left[  \text{\th}\mu-\text{\th}^{\prime}%
\rho+\text{\dh}\pi+\overline{\text{\dh}}\overline{\pi}-\frac{1}{2}%
(\tau-\overline{\pi})(\overline{\tau}-\pi)+\frac{1}{2}\left(  \rho
+\overline{\rho}\right)  \left(  \mu+\overline{\mu}\right)  \right]  $

$\Lambda+\frac{1}{4}\left[  \text{\th}^{\prime}\rho-\text{\th}\mu+\text{\dh
}\pi+\overline{\text{\dh}}\overline{\pi}+\frac{1}{2}\left(  \rho
+\overline{\rho}\right)  \left(  \mu+\overline{\mu}\right)  +\frac{1}{2}%
(\tau-\overline{\pi})(\overline{\tau}-\pi)\right]  .\medskip$

To avoid misunderstandings, we point out that \textit{all }GHP scalars could
be described as 'properly conformally weighted', \textit{provided that the
conformal factor is constant, i.e. for an HV or a KV}. However, when we use
the phrase 'properly conformally weighted' we will be referring to CKVs, for
which the conformal factor is a non-constant function.

Key to our approach in the GHP formalism is the following

\begin{theorem}
In a preferred GHP tetrad relative to a (C)KV $\xi,$ with $\varphi$ the
conformal factor,
\begin{equation}
\left[  \text{\th},\text{\L}_{_{\xi}}\right]  \eta=\frac{\varphi}{2}\text{\th
}\eta+\frac{p+q}{4}\eta\text{\th}\varphi\label{A11}%
\end{equation}
and its companion equations hold for any (p,q)-weighted scalar. Conversely, if
Eqs.(\ref{A11}) (including companions) hold for four functionally independent
(0,0)-weighted quantities and for the same function $\varphi$ throughout, then
$\xi$ is a (C)KV with conformal factor $\varphi$, and the GHP tetrad is
necessarily preferred relative to $\xi$.
\end{theorem}

When attempting to find preferred null directions relative to a number of
(C)KV, it is often very helpful to look for intrinsically defined null
directions such as the principal null directions of the Weyl tensor. Like any
intrinsic quantity, these are fixed in terms of the Riemann tensor and its
covariant derivatives to whichever order is necessary. Such an intrinsic GHP
tetrad exists for most spacetimes but not for all. In practice, unless the
spacetime is conformally flat, the $\mathbf{l}$ - direction of the null tetrad
can be taken to be a principal null direction (p.n.d.) of the Weyl tensor; it
is then intrinsic by definition. If there is a second p.n.d. it can serve as
the $\mathbf{n}$ - direction, giving us an intrinsic GHP tetrad. It was shown
in Ref.\cite{EL2} that an intrinsic GHP tetrad is preferred relative to
\textit{any} KV. In fact, a close look at the proof shows that when the
tetrad's null directions are principal null directions of the Weyl tensor (or
if they are otherwise intrinsically fixed in a proper conformally weighted
manner), this is so even for an HV or a CKV. In type N, in the case of
vanishing spin-coefficient $\rho$, it may still be possible to fix the
$\mathbf{n}$ - direction in terms of the Riemann tensor but such an intrinsic
$\mathbf{n}$ - direction is not necessarily preferred relative to a CKV.

Worse perhaps, there are metrics for which it is not possible to fix the
second null direction intrinsically, and we then have null direction isotropy.
But it should be remembered that relative to a (C)KV, there always does exist
a \emph{preferred} $\mathbf{n}$ - direction irrespective of whether there
exists an intrinsic one. Since we then may not know explicitly what that
preferred null direction is, we have to leave some arbitrariness in the GHP
operators of the GHP-Lie commutator equations (\ref{A11}). The GHP tetrad used
is now defined only up to an unknown null rotation parameter. The equations
become a bit more involved; nevertheless it is usually still quite easy to
solve them. Along with the Lie derivatives of the coordinates, i.e. the
components of the (C)KV, we now obtain the Lie derivative of the null rotation
parameter. If that turns out to be zero then, in hindsight, the null rotation
parameter could have been taken to be zero (since a preferred GHP tetrad is
defined only up to Eqs.(\ref{A8})).

When we are unable to fix the $\mathbf{n}$ - direction intrinsically, an
alternative method is to use the formalism of Machado and Vickers \cite{MV1};
but then it becomes necessary to define a new generalized Lie derivative
operator and to determine the generalizations of Lemmas 4 and 7 and of Theorem
8 to that formalism.

\section{Preferred tetrads and the NP-Lie commutator equations}

While the GHP approach to finding (C)KV is our favourite, the NP formalism is
still much more popular than the GHP one. For this reason we now adapt our
technique to the NP formalism.

Let us start with a tetrad where the null directions have been fixed somehow.
There still remains the two-dimensional gauge freedom
\begin{equation}
\widehat{l}^{^{\mu}}=A^{2}l^{^{\mu}},\widehat{n}^{^{\mu}}=A^{-2}n^{^{\mu}%
},\widehat{m}^{^{\mu}}=\exp\left(  i\alpha_{_{0}}\right)  m^{^{\mu}}.
\label{A12}%
\end{equation}
To obtain a specific gauge and hence a specific tetrad, we need to fix the
gauge factor $A\exp\left(  \frac{i}{2}\alpha_{_{0}}\right)  .$\pagebreak 

\textbf{Definition 2}.\medskip

We define a \emph{preferred gauge relative to a vector }$\xi$\ to be one for
which
\begin{equation}
\text{\L}_{_{\xi}}\eta=\pounds_{_{\xi}}\eta\label{A13}%
\end{equation}
for any quantity $\eta$.

\medskip

\begin{lemma}
There always exists a preferred gauge relative to a vector $\xi$.
\end{lemma}

\begin{proof}
For (0,0) - quantities, Eq.(\ref{A13}) holds in any gauge. Since any other
quantity of interest here is obtained by contracting a tensor with the tetrad
it suffices to show that there is a gauge for which
\begin{equation}
\text{\L}_{_{\xi}}l^{^{\mu}}=\pounds_{_{\xi}}l^{^{\mu}} \label{A14}%
\end{equation}
and companions hold.

>From Eqs.(\ref{A1}),(\ref{A3}) and \ref{A4} we have%

\begin{align}
\text{\L}_{_{\xi}}l^{^{\mu}}  &  =\pounds_{_{\xi}}l^{^{\mu}}-\xi^{^{\nu}%
}\left(  \zeta_{_{\nu}}+\overline{\zeta}_{_{\nu}}\right)  l^{^{\mu}}%
-\frac{1}{2}\left(  \mathcal{P}-\mathcal{P}^{\prime}\right)  l^{^{\mu}%
}\label{A15}\\
\text{\L}_{_{\xi}}n^{^{\mu}}  &  =\pounds_{_{\xi}}n^{^{\mu}}+\xi^{^{\nu}%
}\left(  \zeta_{_{\nu}}+\overline{\zeta}_{_{\nu}}\right)  n^{^{\mu}}%
+\frac{1}{2}\left(  \mathcal{P}-\mathcal{P}^{\prime}\right)  n^{^{\mu}%
}\tag{A15$^\prime$}\\
\text{\L}_{_{\xi}}m^{^{\mu}}  &  =\pounds_{_{\xi}}m^{^{\mu}}-\xi^{^{\nu}%
}\left(  \zeta_{_{\nu}}-\overline{\zeta}_{_{\nu}}\right)  m^{^{\mu}}%
-\frac{1}{2}\left(  \mathcal{P}^{\ast}-\mathcal{P}^{\prime\ast}\right)
m^{^{\mu}}. \tag{A15$^\ast$}%
\end{align}
It is not hard to show that if we choose $A$ and $\alpha_{_{0}}$ such that
\begin{align}
2\pounds_{_{\xi}}\ln A  &  =-\xi^{^{\nu}}\left(  \zeta_{_{\nu}}+\overline
{\zeta}_{_{\nu}}\right)  -\frac{1}{2}\left(  \mathcal{P}-\mathcal{P}^{\prime
}\right) \nonumber\\
i\exp\left(  i\alpha_{_{0}}\right)  \pounds_{_{\xi}}\alpha_{_{0}}  &
=-\xi^{^{\nu}}\left(  \zeta_{_{\nu}}-\overline{\zeta}_{_{\nu}}\right)
-\frac{1}{2}\left(  \mathcal{P}^{\ast}-\mathcal{P}^{\prime\ast}\right)  ,
\label{A16}%
\end{align}
something we can clearly always do, then in the new gauge given by
Eqs.(\ref{A12}), Eqs.(\ref{A14}) (including companions) hold.\medskip\medskip\-
\end{proof}

Note that a preferred gauge is not unique. It is determined only up to boost -
rotations whose parameters satisfy
\begin{equation}
\pounds_{_{\xi}}A=0=\pounds_{_{\xi}}\alpha_{_{0}}. \label{A18}%
\end{equation}

We now wish to combine preferred null directions (i.e., GHP tetrads) with
preferred gauges. Naturally, we make the following\medskip

\medskip

\negthinspace\textbf{Definition 3.\medskip}

A \emph{preferred tetrad relative to a vector }$\xi$\emph{\ }is one for which
both the null directions and the gauge are preferred relative to $\xi$.

\medskip

We are now in a position to prove the following lemmas.

\begin{lemma}
A preferred tetrad relative to a vector $\xi$ always exists.
\end{lemma}

\begin{proof}
We showed in Ref.\cite{EL1} that preferred null directions relative to a
vector $\xi$ always exist and in Lemma 9 that a preferred gauge relative to
$\xi$ always exists as well.
\end{proof}

\begin{lemma}
If the tetrad is Lie recurrent relative to a vector $\xi$ and with associated
factor $\varphi$ then $\xi$ is a (C)KV with $\varphi$ the associated conformal
factor, and the tetrad is necessarily preferred relative to $\xi$.
\end{lemma}

\begin{proof}
Let us first note that if a tetrad is Lie recurrent then the corresponding GHP
tetrad, given by Eq.(\ref{A12}) for an arbitrary gauge factor, satisfies
\begin{equation}
\text{\L}_{_{\xi}}\widehat{l}^{^{\mu}}=-\frac{1}{2}\varphi\widehat{l}^{^{\mu}%
}, \label{A21}%
\end{equation}
with similar equations for its companions. This is shown by basically doing
the reverse of what was done in the proof of Lemma 9. By Lemma 1 and
Eqs.(\ref{A21}) (including companions), $\xi$ is a (C)KV with conformal factor
$\varphi$, and the null directions are preferred. Since, in the original
gauge, \L$_{_{\xi}}$ becomes $\pounds_{_{\xi}}$, this gauge is preferred, and
hence so is the tetrad.\-
\end{proof}

\begin{lemma}
If $\xi$ is a (C)KV with conformal factor $\varphi$ then the tetrad is Lie
recurrent relative to $\xi$ and with associated factor $\varphi$ iff it is
preferred relative to $\xi$.
\end{lemma}

\begin{proof}
Suppose that $\xi$ is a (C)KV. If we assume that the tetrad is preferred
relative to $\xi$ it follows from Lemma 1 that Eq.(\ref{A21}) and its
companions are satisfied. In the given preferred gauge \L$_{_{\xi}}$ becomes
$\pounds_{_{\xi}}$, showing that the tetrad is Lie recurrent.

Conversely, if we assume that the tetrad is Lie recurrent, then by Lemma 11 it
is necessarily preferred.
\end{proof}

\begin{lemma}
If the tetrad is preferred relative to a vector $\xi$ then it is Lie recurrent
relative to $\xi$ with associated factor $\varphi$ iff $\xi$ is a (C)KV with
conformal factor $\varphi$.
\end{lemma}

\begin{proof}
Suppose the tetrad is preferred relative to a vector $\xi$. Then the
corresponding GHP tetrad is preferred. If the tetrad is Lie recurrent then, as
in the proof of Lemma 11, the corresponding GHP tetrad is \emph{GLR}. It
follows from Lemma 3 that $\xi$ is a (C)KV.

Conversely, if $\xi$ is a (C)KV then, since the tetrad is preferred, it
follows also from Lemma 3 that the GHP tetrad is \emph{GLR}. In the preferred
gauge this means that the given tetrad is Lie recurrent.
\end{proof}

\textbf{Definition 4.\medskip}

The \emph{NP scalars with respect to an NP tetrad} are the NP spin
coefficients, all the Riemann tensor tetrad components, together with all
combinations of these and their NP derivatives.\medskip

Next, we come to the main results of this paper, namely the NP versions of
Lemma 7 and Theorem 8. We first show

\begin{lemma}
(a) Given an HV or a KV $\xi$, all combinations of the tetrad, the NP scalars,
and the tetrad components of $\xi$\ itself are Lie recurrent with respect to
$\xi$ provided that the tetrad is preferred relative to $\xi$.\medskip

(b) Given a (C)KV $\xi$, all properly conformally weighted combinations of the
tetrad, the NP scalars, and the tetrad components of $\xi$\ itself are Lie
recurrent with respect to $\xi$ provided that the tetrad is preferred relative
to $\xi$.
\end{lemma}

\begin{proof}
The proofs of (a) and (b) follow immediately from Lemmas 4 and 7 respectively
since, in a preferred gauge, \L$_{_{\xi}}=\pounds_{_{\xi}}.$\medskip\thinspace
Strictly speaking, these proofs only apply to combinations which are of proper
GHP weight; but we can extend the proofs, by direct calculation, to all
combinations of the tetrad and NP scalars.
\end{proof}

For reference, we will quote all conformally weighted NP scalars. They are the
conformally weighted GHP scalars together with $\varepsilon-\overline
{\varepsilon},$ $\gamma-\overline{\gamma},$ $\alpha+\overline{\beta}$ and all
conformally weighted combinations thereof.

\begin{theorem}
In a preferred tetrad relative to a (C)KV $\xi,$ with $\varphi$ the conformal
factor,
\begin{equation}
\left[  D,\pounds_{_{\xi}}\right]  \eta=\frac{\varphi}{2}D\eta\label{A22}%
\end{equation}
and its companion equations hold for any scalar quantity $\eta$. Conversely,
if Eqs.(\ref{A22}) (including companions) hold for four functionally
independent quantities and for the same function $\varphi$ throughout, then
$\xi$ is a (C)KV with conformal factor $\varphi$, and the tetrad is
necessarily preferred relative to $\xi$.
\end{theorem}

\begin{proof}
Suppose that $\xi$ is a (C)KV and that the tetrad is preferred relative to
$\xi$. Since the corresponding GHP tetrad is preferred relative to $\xi$, it
follows from Theorem 8 that
\begin{equation}
\left[  \text{\th},\text{\L}_{_{\xi}}\right]  \eta=\frac{\varphi}{2}\text{\th
}\eta\label{A23}%
\end{equation}
and companion equations hold. Since the gauge is also preferred relative to
$\xi$, Eq.(\ref{A23}) reduces to Eq.(\ref{A22}) for the given preferred
tetrad, and similarly for the companion equations.

Conversely, if Eq.(\ref{A22}) and its companions hold in some tetrad for four
functionally independent quantities then, for the corresponding GHP tetrad,
Eq.(\ref{A23}) and its companions hold. But, by Theorem 8, this means that
$\xi$ is a (C)KV and the GHP tetrad is necessarily preferred. Since the gauge
is preferred iff \L$_{_{\xi}}=\pounds_{_{\xi}}$ it follows that the tetrad is preferred.
\end{proof}

We have noted at the end of the last section that in order to find preferred
tetrad directions (i.e. preferred GHP tetrads) it is often useful to look for
intrinsically defined null directions, which can be shown to be preferred for
all KV. Furthermore, it can be shown that intrinsically defined null
directions which are defined in a proper conformally weighted manner (usually
in terms of the Weyl tensor) are preferred with respect to all (C)KV. The
principal null directions are particularly useful in this regard.

Similarly, in order to find a preferred gauge, it is often useful to determine
if an intrinsic restriction determines the gauge; such a gauge will be
preferred for all KV. In order to find a gauge which is preferred with respect
to all (C)KVs, it is useful to look at a gauge defined both intrinsically and
in a proper conformally weighted manner (which often, but not necessarily,
means defined with respect to the Weyl tensor). Such a gauge can be shown to
be preferred as follows.

We must show that, for such a gauge, \L$_{_{\xi}}\eta=\pounds_{_{\xi}}\eta
$\ for all GHP scalars $\eta$. It suffices to show this for one ($p$,$q$) -
weighted quantity $R$\ with $p\neq\pm q$. Suppose that the gauge is fixed by
the condition $R=1,$ where $R$\ is intrinsic. For this to be a properly
conformally weighted condition, the conformal weight of $R$\ must be zero, i.e
the $w$\ - weight of $R$\ must vanish. But then \L$_{_{\xi}}R=0=\pounds
_{_{\xi}}R$\textbf{.}

\section{Metrics in Canonical Tetrads}

Because of Theorem 15, our approach to finding (C)KVs is as follows. In a
preferred tetrad relative to the (C)KV to be found we apply Eq.(\ref{A22}) and
its companions to the four coordinates $x^{i}$ and solve these equations for
$\pounds_{_{\xi}}x^{i}$ ($i=1,2,3,4$), thereby obtaining the components of the
(C)KV. If we have succeeded in choosing a tetrad which is preferred relative
to a number of different (C)KVs, then this solution will yield all such (C)KVs.

In order to illustrate simply how our results can be exploited in practice we
will now restrict our consideration to metrics which have been constructed
following standard routines via an NP tetrad. In principle, we know that we
can investigate the symmetries of such metrics via Theorem 15. Obviously, if
the NP tetrad is a preferred tetrad then we just apply the theorem directly.
Unfortunately, the concept of a 'preferred tetrad' is not used explicitly in
the standard NP routines, and so our first task will be to look in some detail
at how NP tetrads are constructed, and how they relate to our concept of a
'preferred tetrad'. Fortunately, the manner in which NP tetrads are often
chosen --- at least partly determined by the Weyl or Ricci (equivalently
energy-momentum) tensor and their covariant derivatives --- enables us to find
these relations explicitly.

A standard NP approach to determining a metric, whose Weyl tensor is required
to be of a particular Petrov type, involves adopting the canonical null tetrad
associated with that particular Petrov type as the NP tetrad for the
calculations. In order for us to exploit such results, we need to know how
close the canonical null tetrad of each Petrov type is to a preferred tetrad.

So, in the following practical applications to specific metrics already quoted
in canonical NP tetrads, our approach will be first to determine the Petrov
type which will tell us to what extent the NP canonical tetrad is related to a
preferred tetrad. Once we have transformed the given NP canonical tetrad into
a preferred tetrad by a suitable transformation, then we can apply Theorem 15
and use the Lie-NP commutators to determine KVs, HVs and CKVs appropriately.
However, in order for our calculations to be as efficient as possible, it is
important that we also fully exploit Lemma 14. Since the coordinates to which
we apply the Lie-NP commutators are frequently defined in terms of the NP
scalars, exploiting the simple recurrence properties of this lemma can
considerably simplify the calculations involving the commutators.

\subsection{\textbf{PETROV TYPE I} with a canonical tetrad}

A canonical null tetrad$^{\text{F}}$\footnote{It should be noted that there
are alternatives for a canonical tetrad of type I spaces, e.g. $\Psi_{_{0}%
}=0=\Psi_{_{4}}$, $\Psi_{_{1}}=\Psi_{_{3}}$, $\Psi_{_{2}}$ $\neq0$; but we
have used the alternative employed in the IC program\cite{Karl}.} for Petrov
type I is defined by

$\Psi_{_{1}}=0=\Psi_{_{3}}$, $\Psi_{_{0}}$ $=\Psi_{_{4}}$, $\Psi_{_{2}}$ $\neq0.$

Since the tetrad directions as well as the gauge are defined by the Weyl
tensor in a proper conformally weighted manner, the tetrad is easily seen to
be preferred for all (C)KVs. So we can immediately apply Theorem 15. Being
able to decide whether the coordinates of the metric under investigation are
Lie recurrent is important for carrying out our calculations with the NP-Lie
commutators in the most efficient manner. Lemma 14 gives us the crucial result
for the scalars associated with the preferred tetrad. In this case, the
canonical tetrad is preferred. Therefore, we can take the conclusions of Lemma
14 completely across to the canonical NP scalars associated with the canonical
NP tetrad; hence those coordinates which have been defined from the canonical
NP scalars of the canonical NP tetrad are Lie recurrent. We have established
the following

\begin{theorem}
If a given metric is of Petrov I and if an NP tetrad for this metric is in the
above canonical form, then all (C)KVs can be found by applying $[D,\pounds
_{_{\xi}}]=\frac{1}{2}\varphi D$ and its companion equations to the four
coordinates $x^{i}$, and solving for $\pounds_{_{\xi}}x^{i}$ and the conformal
function $\varphi$. Furthermore, (a) all proper conformally weighted
combinations of the NP scalars are Lie recurrent relative to each (C)KV, (b)
\emph{all} combinations of the NP scalars are Lie recurrent relative to each
KV and HV.
\end{theorem}

\smallskip

\textbf{Example 1.\medskip}

\smallskip

The Bell-Szekeres\cite{BS} metric, in coordinates $\left(  u,v,x,y\right)  $
is given by
\begin{equation}
ds^{2}=2\exp\left(  \frac{U}{2}\left(  a^{2}-1\right)  \right)  dudv-2\exp
\left(  U\left(  1-a\right)  \right)  dx^{2}-2\exp\left(  U\left(  1+a\right)
\right)  dy^{2}, \label{E1}%
\end{equation}
where $U=\ln\left(  u+v\right)  $. For a tetrad we may take
\begin{align}
l^{^{\alpha}}  &  =\left(  \exp\left(  \frac{U}{4}\left(  1-a^{2}\right)
\right)  ,0,0,0\right) \nonumber\\
n^{^{\alpha}}  &  =\left(  0,\exp\left(  \frac{U}{4}\left(  1-a^{2}\right)
\right)  ,0,0\right) \nonumber\\
m^{^{\alpha}}  &  =\left(  0,0,\frac{1}{2}\exp\left(  \frac{U}{2}\left(
a-1\right)  \right)  ,\frac{i}{2}\exp\left(  -\frac{U}{2}\left(  a+1\right)
\right)  \right)  . \label{E2}%
\end{align}
Therefore,%

\begin{align}
Du  &  =\exp\left(  \frac{U}{4}\left(  1-a^{2}\right)  \right)  \hspace
{1.5cm}Dv=0\hspace{1.5cm}Dx=0\hspace{1.5cm}Dy=0\nonumber\\
\Delta u  &  =0\hspace{1.5cm}\Delta v=\exp\left(  \frac{U}{4}\left(
1-a^{2}\right)  \right)  \hspace{1.5cm}\Delta x=0\hspace{1.5cm}\Delta
y=0\nonumber\\
\delta u  &  =0\hspace{0.6cm}\delta v=0\hspace{0.6cm}\delta x=\frac{1}{2}%
\exp\left(  \frac{U}{2}\left(  a-1\right)  \right)  \hspace{0.55cm}\delta
y=\frac{i}{2}\exp\left(  -\frac{U}{2}\left(  a+1\right)  \right)  . \label{E3}%
\end{align}
The spin coefficients are
\begin{align}
\kappa &  =\nu=\tau=\pi=\alpha=\beta=0\nonumber\\
\lambda &  =-\frac{a}{2}DU=-\sigma\;\;\hspace{1cm}\mu=\frac{1}{2}%
DU=-\rho\;\;\hspace{1cm}\;\varepsilon=\frac{a^{2}-1}{8}DU=-\gamma, \label{E4}%
\end{align}
where
\begin{equation}
DU=\frac{1}{u+v}\exp\left(  \frac{U}{4}\left(  1-a^{2}\right)  \right)
=\exp\left(  -\frac{U}{4}\left(  3+a^{2}\right)  \right)  . \label{E5}%
\end{equation}
The tetrad components of the Riemann tensor are
\begin{align}
\Phi_{_{00}}  &  =\Phi_{_{01}}=\Phi_{_{02}}=\Phi_{_{12}}=\Phi_{_{11}}%
=\Phi_{_{22}}=\Lambda=\Psi_{_{1}}=\Psi_{_{3}}=0,\nonumber\\
\Psi_{_{0}}  &  =\frac{a}{4}\left(  1-a^{2}\right)  \exp\left(  -\frac{U}%
{2}\left(  3+a^{2}\right)  \right)  =\Psi_{_{4}},\nonumber\\
\Psi_{2}  &  =\frac{a^{2}-1}{4}\exp\left(  -\frac{U}{2}\left(  3+a^{2}\right)
\right)  . \label{E6}%
\end{align}
We assume that $a\neq0$ (otherwise the metric is type D). We also assume that
$a\neq\pm1$ since otherwise the metric is flat. The metric is then vacuum, of
Petrov type I. The tetrad is determined completely by the Weyl tensor, and the
gauges are defined via properly conformally weighted conditions ($\Psi_{_{0}%
}=\Psi_{_{4}}$). We can therefore apply Theorem 16. But since the metric is
vacuum we already know\cite{Hall} that no CKVs exist, so we need only consider
$\varphi$ constant and use part (a).

Although we can use Theorem 16 (a) to determine the Lie recurrence relations
for the NP scalars, we are not able to find simple explicit recurrence
relations for any of the coordinates. However, if, instead, we focus on $u+v$
and $u-v$, we can use the result that $\pounds_{_{\xi}}\Psi_{_{2}}=$
$-\varphi\Psi_{_{2}}$ to get $\pounds_{_{\xi}}\left(  u+v\right)
=\frac{2\varphi}{3+a^{2}}\left(  u+v\right)  .$ The commutator equations on
$\left(  u-v\right)  $ integrate to give
\begin{equation}
\pounds_{_{\xi}}\left(  u-v\right)  =\frac{2\varphi}{3+a^{2}}\left(
u-v\right)  +c_{_{1}}, \label{E7}%
\end{equation}
where $c_{_{1}}$ is an integration constant. Hence,
\begin{align}
\pounds_{_{\xi}}u  &  =\frac{c_{1}}{2}+\frac{2\varphi}{3+a^{2}}u\nonumber\\
\pounds_{_{\xi}}v  &  =-\frac{c_{1}}{2}+\frac{2\varphi}{3+a^{2}}v. \label{E8}%
\end{align}
The NP-Lie commutator equations on $x$ and $y$ integrate to give,
respectively,
\begin{align}
\pounds_{_{\xi}}x  &  =\frac{\varphi}{2\left(  3+a^{2}\right)  }\left(
1+a\right)  ^{2}x+c_{_{2}}\nonumber\\
\pounds_{_{\xi}}y  &  =\frac{\varphi}{2\left(  3+a^{2}\right)  }\left(
1-a\right)  ^{2}y+c_{_{3}}. \label{E9}%
\end{align}
Therefore,
\begin{align}
\xi &  =\frac{2\varphi}{3+a^{2}}\left[  \left(  u+v\right)  \frac{\partial
}{\partial\left(  u+v\right)  }+\left(  u-v\right)  \frac{\partial}%
{\partial\left(  u-v\right)  }+\frac{x}{4}\left(  1+a\right)  ^{2}%
\frac{\partial}{\partial x}+\frac{y}{4}\left(  1-a\right)  ^{2}\frac{\partial
}{\partial y}\right] \nonumber\\
&  +c_{_{1}}\frac{\partial}{\partial\left(  u-v\right)  }+c_{_{2}%
}\frac{\partial}{\partial x}+c_{_{3}}\frac{\partial}{\partial y}. \label{E10}%
\end{align}
The KVs are $\frac{\partial}{\partial u}-\frac{\partial}{\partial
v},\frac{\partial}{\partial x},$ and $\frac{\partial}{\partial y}$. The HV is
\begin{equation}
4u\frac{\partial}{\partial u}+4v\frac{\partial}{\partial v}+\left(
1+a\right)  ^{2}x\frac{\partial}{\partial x}+\left(  1-a\right)
^{2}y\frac{\partial}{\partial y}. \label{E11}%
\end{equation}

\subsection{\textbf{PETROV TYPE II} with a canonical tetrad}

A canonical tetrad for Petrov type II is defined$^{\text{F}}$\footnote{It
should be noted that the value of the constant for $\Psi_{_{4}}$ differs for
different references; we have chosen here, and in the other Petrov types to
follow, the value used in the IC program\cite{Karl}.} by $\Psi_{_{4}}%
=1,\Psi_{_{2}}$ $\neq0,$ all other $\Psi$ equal to zero.

In Petrov type II the canonical tetrads are fixed completely by the Weyl
tensor, and it is easy to see that such a tetrad is a preferred tetrad for any
KV that may exist in the spacetime. Therefore, we can immediately apply
Theorem 15 to investigate KVs. However, the above canonical tetrad will not be
preferred for CKVs, or even for an HV, since the standard gauge specialization
($\Psi_{_{4}}=1$) is not a proper conformally weighted condition.

Therefore, in Petrov type II, although a canonical tetrad is sufficient to
investigate KVs, to become a preferred tetrad with reference to an HV or a
CKV, it will require a gauge transformation.\textbf{\ }To achieve this we
simply introduce a gauge parameter $A\exp\left(  \frac{i}{2}\alpha_{_{0}%
}\right)  =(\Psi_{_{4}}/\Psi_{_{2}})^{^{\frac{1}{4}}}$, so that \textit{in the
transformed tetrad, }$\Psi_{_{4}}=\Psi_{_{2}}$, and this gauge specialization
is now preferred. Therefore, for \textit{the transformed tetrad} we can apply
Theorem 15 and Lemma{\Large \ }14 directly.

\begin{theorem}
If a given metric is of Petrov II and if an NP tetrad for this metric is in
the above canonical form then, after a gauge transformation $A\exp\left(
\frac{i}{2}\alpha_{_{0}}\right)  =$ $\left(  \frac{\Psi_{_{4}}}{\Psi_{2}%
}\right)  ^{\frac{1}{4}}$ to a new tetrad $\widehat{l},\widehat{n},\widehat
{m},\widehat{\overline{m}}$, all (C)KVs can be found by applying $[\widehat
{D},\pounds]=\frac{1}{2}\varphi\widehat{D}$ and its companion equations to the
four coordinates $x^{i}$, and solving for $\pounds_{_{\xi}}x^{i}$ and the
conformal function $\varphi$. Furthermore, (a) in the transformed tetrad, all
properly conformally weighted combinations of the NP scalars are Lie recurrent
relative to each (C)KV, (b) in the transformed tetrad, all combinations of the
NP scalars are Lie recurrent relative to each KV and HV.
\end{theorem}

\begin{corollary}
The \emph{KVs only} can be found by applying $[D,\pounds]=0$ and its companion
equations to the four coordinates $x^{i}$, and solving for $\pounds_{_{\xi}%
}x^{i}$. Furthermore, in the original tetrad, all combinations of NP scalars
are Lie derived relative to each KV.\medskip
\end{corollary}

Note that the statement of this theorem is a bit more general than is required
for a canonical tetrad of Petrov type II. In fact, it can apply to a tetrad
with any gauge choice, not just for the canonical choice $\Psi_{_{4}%
}=1.\medskip$\textbf{\ }

\smallskip

\textbf{Example 2.} \medskip

The Steele metric\cite{Steele} considered by Koutras and Skea\cite{KS} is
given by%

\begin{equation}
ds^{2}=r\ln rdu^{2}+2rdxdu-r^{^{-\frac{1}{2}}}\left(  dr^{2}+dy^{2}\right)  .
\label{B1}%
\end{equation}
In coordinates $\left(  u,r,x,y\right)  $ the tetrad vectors can be given by%

\begin{align}
l^{^{\alpha}}  &  =\left(  2\sqrt{2}r^{^{\frac{1}{4}}},0,-\sqrt{2}%
r^{^{\frac{1}{4}}}\ln r,0\right) \nonumber\\
n^{^{\alpha}}  &  =\left(  0,0,\frac{1}{2\sqrt{2}}r^{^{-\frac{5}{4}}},0\right)
\nonumber\\
m^{^{\alpha}}  &  =\left(  0,-\frac{1}{\sqrt{2}}r^{^{\frac{1}{4}}}%
,0,-\frac{i}{\sqrt{2}}r^{^{\frac{1}{4}}}\right)  . \label{B2}%
\end{align}
>From Eq.(\ref{B2}) we obtain
\begin{equation}%
\begin{array}
[c]{llll}%
Du=2\sqrt{2}r^{^{\frac{1}{4}}} & Dr=0 & Dx=-\sqrt{2}r^{^{\frac{1}{4}}}\ln r &
Dy=0\\
\triangle u=0 & \triangle r=0 & \triangle x=\frac{1}{2\sqrt{2}}r^{^{-\frac{5}%
{4}}} & \triangle y=0\\
\delta u=0 & \delta r=-\frac{1}{\sqrt{2}}r^{^{\frac{1}{4}}} & \delta x=0 &
\delta y=-\frac{i}{\sqrt{2}}r^{^{\frac{1}{4}}}.
\end{array}
\label{B3}%
\end{equation}

The spin-coefficients are given by
\begin{align}
\nu &  =\rho=\sigma=\lambda=\mu=\varepsilon=\gamma=0\nonumber\\
\tau &  =-\overline{\pi}=\frac{1}{2\sqrt{2}}r^{^{-\frac{3}{4}}},\qquad
\kappa=2\sqrt{2}r^{^{\frac{3}{4}}},\qquad\alpha=-\frac{1}{2\sqrt{2}%
}r^{^{-\frac{3}{4}}}=2\beta. \label{B4}%
\end{align}
and the nonzero components of the Riemann tensor by
\begin{equation}
\Psi_{0}=1,\qquad\Psi_{2}=\frac{1}{8}r^{-^{\frac{3}{2}}}. \label{B5}%
\end{equation}

The metric is vacuum, of Petrov type II. The tetrad is canonical, with the
gauge fixed by $\Psi_{_{0}}=1$; therefore, we apply the gauge transformation
$A\exp\left(  \frac{i}{2}\alpha_{_{0}}\right)  =(\frac{\Psi_{2}}{\Psi_{0}%
})^{^{\frac{1}{4}}}$ and use Theorem 17 (slightly modified because it is
$\Psi_{0\text{ }}$, not $\Psi_{4\text{ }}$, which equals one) to investigate
all (C)KVs. Since the metric is vacuum we already know that no CKVs\cite{Hall}
exist; so we can apply Theorem 17 to find the HV using $\varphi$ = constant
and statement (b).

Since $\Psi_{_{2}}$ is Lie recurrent according to $\pounds_{_{\xi}}\Psi_{_{2}%
}=-\varphi\Psi_{_{2}}$, we can deduce that the coordinate $r$ is Lie recurrent
according to $\pounds_{_{\xi}}r=\frac{2}{3}\varphi$ $r$. Therefore, we first
substitute $r$ into the commutators and find that the latter are identically satisfied.

Applying the NP-Lie commutator equations to the remaining coordinates
\textit{in the transformed tetrad} and solving these equations we find that
\begin{align}
\varphi &  =\text{constant},\quad\pounds_{_{\xi}}u=\frac{1}{6}\varphi
u+k,\quad\nonumber\\
\quad\pounds_{_{\xi}}x  &  =\frac{1}{6}\varphi x-\frac{1}{3}\varphi
u+b,\qquad\pounds_{_{\xi}}y=\frac{2}{3}\varphi y+d \label{B6}%
\end{align}
where $k$, $b$, $d$ are integration constants. Therefore,
\begin{equation}
\xi=\left(  \frac{1}{6}\varphi u+k,\frac{2}{3}\varphi r,\frac{1}{6}\varphi
x-\frac{1}{3}\varphi u+b,\frac{2}{3}\varphi y+d\right)  . \label{B7}%
\end{equation}
Putting, in turn, one of the integration constants equal to one and the
remaining ones equal to zero we get the HV
\begin{equation}
\xi_{_{1}}=u\frac{\partial}{\partial u}+4r\frac{\partial}{\partial r}+\left(
x-2u\right)  \frac{\partial}{\partial x}+4y\frac{\partial}{\partial y}
\label{B8}%
\end{equation}
and the three KVs
\begin{equation}
\xi_{_{2}}=\frac{\partial}{\partial u},\qquad\xi_{_{3}}=\frac{\partial
}{\partial x},\qquad\xi_{4}=\frac{\partial}{\partial y}. \label{B9}%
\end{equation}
Of course, the KV were obvious since the metric is independent of the $u$, $x
$, and\linebreak $y$ - coordinates. There are no proper CKVs.

\subsection{\textbf{PETROV TYPE III} with a canonical tetrad}

In Petrov type III the canonical tetrad is fixed completely by the Weyl tensor:

$\Psi_{_{3}}=1$, all other $\Psi$ zero. Although this tetrad is preferred for
any KVs that may exist in the spacetime, it is not necessarily preferred for
an HV or for CKVs. Therefore, in Petrov type III, although a canonical tetrad
is sufficient to investigate KVs, in order to become a preferred tetrad with
respect to a CKV, it will require a gauge transformation. Unfortunately, there
are no other $\Psi$ components with which we can make a suitable gauge
transformation (as was possible for Petrov II). So we may have to introduce
explicitly an arbitrary gauge parameter\textbf{\ }which transforms the
canonical tetrad to a preferred tetrad with respect to which we can then apply
Theorem 15; we will find the value(s) for this parameter simultaneously while
solving for the (C)KVs. In fact, because the condition $\Psi_{_{3}}%
-\overline{\Psi_{_{3}}}=0$ is properly conformally weighted, and it is only
the condition $\Psi_{_{3}}+\overline{\Psi_{_{3}}}=2$ which is not properly
conformally weighted, we will require only one arbitrary gauge parameter,
namely a boost parameter $A$, to become a preferred tetrad with respect to a CKV.

Lemma 14 gives us the crucial recurrence results for the NP scalars associated
with the preferred tetrad. Unfortunately, for the CKV case, the canonical
tetrads are not preferred and so we cannot take the conclusion of Lemma 14
completely across to the NP scalars associated with the canonical tetrad; but
we can use Lemma 14 for those NP scalars \emph{which are identical in both
canonical and preferred tetrads}, i.e., NP scalars which are \emph{independent
of boost} (i.e. those whose boost weight is zero), and which have proper
conformal weight.

\begin{theorem}
If a given metric is of Petrov III and if an NP tetrad for this metric is in
the above canonical form, then after a boost with a real parameter $A$ to a
new tetrad $\widehat{l},\widehat{n},\widehat{m},\widehat{\overline{m}}$, all
(C)KVs can be found by applying $[\widehat{D},\pounds]=\frac{1}{2}%
\varphi\widehat{D}$ and its companion equations to the four coordinates
$x^{i}$, and solving for $\pounds x^{i}$, for $A$, and for the conformal
function $\varphi$. Furthermore, (a) all boost- and properly conformally
weighted combinations of the NP scalars are Lie recurrent relative to each
(C)KV, (b) all boost-invariant combinations of the NP scalars are Lie
recurrent relative to an HV or a KV.
\end{theorem}

\begin{corollary}
The \emph{KVs only} can be found by putting $A=1,\varphi=0$, and solving the
commutator equations for\textbf{\ }$\pounds_{_{\xi}}x^{i}$. Furthermore, all
combinations of the NP scalars are Lie derived relative to each KV.
\end{corollary}

\smallskip

\subsection{\textbf{PETROV TYPE D }with a canonical tetrad}

In Petrov type D, a canonical tetrad is defined by $\Psi_{_{2}}\neq0$, all
other $\Psi$ zero. It is clear that such a tetrad has preferred null
directions, but all the gauge freedom remains arbitrary. Often this freedom is
fixed explicitly in the process of determining the metric, either by reference
to the Riemann tensor or the spin coefficients. However, it is important to
realize that sometimes it is very difficult to fix the tetrad completely
because of the complexity of the calculations, and also sometimes there is
some unavoidable residual freedom due to isotropy which makes it impossible to
fix the tetrad using the Riemann tensor and its derivatives. As in Type III,
there are no more nonzero Weyl tensor components to enable us to fix the
tetrad (but of course, in general, there could be Ricci tensor components, and
derivatives of the Weyl and Ricci tetrad components).

If the gauge of the canonical tetrad has been fixed with respect to the
Riemann tensor or its derivatives, then the tetrad is sufficient to
investigate KVs. On the other hand, if the gauge has still some freedom, or is
not fixed in a properly conformally weighted manner, or indeed if precisely
how it is fixed is not immediately clear, we will still need to make a gauge
transformation from the canonical to a preferred tetrad; arbitrary gauge
parameters will have to be introduced which will be solved for, simultaneously
with the (C)KVs.

For the CKV case, we cannot use Lemma 14 for \textit{all} canonical NP
scalars, but we can apply it to those scalars which are identical in both the
canonical and preferred tetrad, i.e., scalars which are gauge invariant (whose
spin and boost weights are both zero) and which have proper conformal weight.

\begin{theorem}
If a given metric is of Petrov D and if an NP tetrad for this metric is in the
above canonical form then, after a gauge transformation with complex parameter
$A\exp\left(  \frac{i}{2}\alpha_{_{0}}\right)  $ to a new tetrad $\widehat
{l},\widehat{n},\widehat{m},\widehat{\overline{m}}$, all (C)KVs can be found
by applying $[\widehat{D},\pounds]=\frac{1}{2}\varphi\widehat{D}$, and its
companion equations to the four coordinates $x^{i}$, and solving for
$\pounds_{_{\xi}}$ $x^{i}$, the unknown gauge parameters $A$, $\alpha_{_{0}}$,
and the conformal function $\varphi$. Furthermore, (a) all gauge invariant and
properly conformally weighted combinations of the NP scalars are Lie recurrent
relative to each (C)KV, (b) all gauge invariant combinations of the NP scalars
are Lie recurrent relative to an HV or a KV.
\end{theorem}

It will be useful to have explicit examples of gauge invariant NP
scalars,\linebreak e.g., $\rho\mu$, $\pi\tau$, $\kappa\nu$, $\lambda\sigma$,
$\Psi_{_{2}}$, $\Psi_{_{0}}\Psi_{_{4}}$, $\Lambda$, $\Phi_{_{11}}$,
$\Phi_{_{00}}\Phi_{_{22}}$, $\Phi_{_{02}}\Phi_{_{20}}$

and gauge invariant and properly conformally weighted NP scalars,\linebreak
e.g., $\left(  \rho-\overline{\rho}\right)  \left(  \mu-\overline{\mu}\right)
$, $\left(  \tau+\overline{\pi}\right)  \left(  \overline{\tau}+\pi\right)  $,
$\kappa\nu$, $\lambda\sigma$, $\Psi_{_{2}}$, $\Psi_{_{0}}\Psi_{_{4}}%
\smallskip\medskip.$

\medskip\textbf{Example 3.\medskip}

\medskip The Kimura metric\cite{Kimura} considered by Koutras and
Skea\cite{KS}, given by
\begin{equation}
ds^{2}=\frac{r^{2}}{b}dt^{2}-\frac{1}{r^{2}b^{2}}dr^{2}-r^{2}d\theta^{2}%
-r^{2}\sin^{2}\theta d\phi^{2}, \label{A41}%
\end{equation}
is of Petrov type D with a non-zero energy momentum tensor. The metric is
diagonal and, in coordinates $\left(  t,r,\theta,\phi\right)  $, we can
readily construct the tetrad
\begin{align}
l_{_{\alpha}}  &  =\left(  \frac{r}{\sqrt{2b}},-\frac{1}{\sqrt{2}%
rb},0,0\right) \nonumber\\
n_{_{\alpha}}  &  =\left(  \frac{r}{\sqrt{2b}},\frac{1}{\sqrt{2}rb},0,0\right)
\nonumber\\
m_{_{\alpha}}  &  =\left(  0,0,-\frac{r}{\sqrt{2}},-\frac{ir}{\sqrt{2}}%
\sin\theta\right)  , \label{A42}%
\end{align}
or, in contravariant form,\newpage%
\begin{align}
l^{^{\alpha}}  &  =\left(  \frac{\sqrt{b}}{r\sqrt{2}},\frac{rb}{\sqrt{2}%
},0,0\right) \nonumber\\
n^{^{\alpha}}  &  =\left(  \frac{\sqrt{b}}{r\sqrt{2}},-\frac{rb}{\sqrt{2}%
},0,0\right) \nonumber\\
m^{^{\alpha}}  &  =\left(  0,0,\frac{1}{r\sqrt{2}},\frac{i}{r\sqrt{2}%
\sin\theta}\right)  . \label{A43}%
\end{align}
It follows that
\begin{equation}%
\begin{array}
[c]{llll}%
Dt=\frac{\sqrt{b}}{r\sqrt{2}}, & Dr=\frac{rb}{\sqrt{2}}, & D\theta=0, &
D\phi=0\\
\triangle t=\frac{\sqrt{b}}{r\sqrt{2}}, & \triangle r=-\frac{rb}{\sqrt{2}}, &
\triangle\theta=0, & \triangle\phi=0\\
\delta t=0, & \delta r=0, & \delta\theta=\frac{1}{r\sqrt{2}}, & \delta
\phi=\frac{i}{r\sqrt{2}\sin\theta}.
\end{array}
\label{A44}%
\end{equation}

The NP spin coefficients are
\begin{align}
\kappa &  =\sigma=\lambda=\nu=\tau=\pi=0\nonumber\\
\gamma &  =\varepsilon=\frac{b}{2\sqrt{2}},\qquad\rho=\mu=-\frac{b}{\sqrt{2}%
},\qquad\beta=-\overline{\alpha}=\frac{\cot\theta}{2\sqrt{2}r}, \label{A45}%
\end{align}
and
\begin{equation}
\Psi_{_{2}}=-\frac{1}{6r^{2}},\qquad\Phi_{_{11}}=\frac{1}{4r^{2}}%
,\qquad\Lambda=-\frac{b^{2}}{2}+\frac{1}{12r^{2}} \label{A46}%
\end{equation}
are the only nonzero components of the Riemann tensor.

The tetrad is canonical, but it is not immediately obvious how, and to what
extent, the gauge has been fixed; we cannot use the Ricci components and
$\Psi_{_{2}}$ to fix the gauge since all three components are invariant under
gauge. So we apply Theorem 21 to investigate all (C)KVs.

We shall first investigate Lie recurrence relations for all (C)KV
possibilities together, using (a). We note that $\Psi_{_{2}}$ is gauge
invariant and of conformal weight $w=-2$, and therefore $\pounds_{_{\xi}}%
\Psi_{_{2}}=-\varphi\Psi_{_{2}}$. Hence for the coordinate $r$ we have
$\pounds_{_{\xi}}r=\frac{1}{2}\varphi r$. Although there is a second
coordinate $\theta$ explicitly in the metric it does not occur in any gauge
invariant properly conformally weighted scalars, and so we cannot use Theorem
21 to obtain a Lie recurrence relation for $\theta$. (On the other hand,
$\theta$ does occur in the spin coefficients $\alpha$ and $\beta$, but these
coefficients do not have appropriate behaviour for the theorem to apply.)

We therefore insert both a phase factor and a boost into our tetrad, so that
Eq.(\ref{A44}) becomes
\begin{equation}%
\begin{array}
[c]{llll}%
Dt=\frac{A\sqrt{b}}{r\sqrt{2}}, & Dr=\frac{Arb}{\sqrt{2}}, & D\theta=0, &
D\phi=0\\
\triangle t=\frac{\sqrt{b}}{Ar\sqrt{2}}, & \triangle r=-\frac{rb}{A\sqrt{2}%
}, & \triangle\theta=0, & \triangle\phi=0\\
\delta t=0, & \delta r=0, & \delta\theta=\frac{\exp\left(  i\alpha_{_{0}%
}\right)  }{r\sqrt{2}}, & \delta\phi=\frac{i\exp\left(  i\alpha_{_{0}}\right)
}{r\sqrt{2}\sin\theta}.
\end{array}
\label{C1}%
\end{equation}

Turning next to the NP-Lie commutator equations applied to the four
coordinates we obtain, in addition to $\pounds_{_{\xi}}r=\frac{1}{2}\varphi
r,$ the following results.
\begin{equation}
\pounds_{_{\xi}}t=h(r),\qquad\qquad\pounds_{_{\xi}}\theta=f(\phi
),\qquad\pounds_{_{\xi}}\phi=g(\theta,\phi), \label{C2}%
\end{equation}
where these four functions have to satisfy the equations
\begin{align}
\frac{\partial g(\theta,\phi)}{\partial\theta}  &  =-\frac{1}{\sin\theta
}\pounds_{_{\xi}}\alpha_{_{0}}\nonumber\\
\frac{df(\phi)}{d\phi}  &  =\sin\theta\pounds_{_{\xi}}\alpha_{_{0}}\nonumber\\
\frac{\partial g(\theta,\phi)}{\partial\phi}+\cot\theta f(\phi)  &
=0\nonumber\\
\frac{1}{2}r\frac{\partial\varphi}{\partial r}  &  =\frac{\varphi
(t,r,\theta,\phi)}{2}\nonumber\\
A^{-1}\pounds_{_{\xi}}A  &  =r^{2}\sqrt{b}\frac{dh(r)}{dr}\nonumber\\
A^{-1}\pounds_{_{\xi}}A  &  =\frac{1}{2r\sqrt{b}}\frac{\partial\varphi
}{\partial t} \label{C3}%
\end{align}

These equations are readily solved and we find that
\begin{align}
\varphi &  =\varphi(t,r)=r\left(  2l_{_{0}}bt+l_{_{1}}\right) \nonumber\\
\pounds_{_{\xi}}r  &  =\frac{1}{2}r^{2}\left(  2l_{_{0}}bt+l_{_{1}}\right)
\nonumber\\
\pounds_{_{\xi}}t  &  =h\left(  r\right)  =-\frac{l_{_{0}}}{r}+h_{_{0}}%
,\qquad\nonumber\\
\pounds_{_{\xi}}\theta &  =f\left(  \phi\right)  =a_{_{0}}\cos\phi+b_{_{0}%
}\sin\phi\qquad\nonumber\\
\pounds_{_{\xi}}\phi &  =g\left(  \theta,\phi\right)  =c_{_{0}}+\cot
\theta\left(  -a_{_{0}}\sin\phi+b_{_{0}}\cos\phi\right)  \label{C4}%
\end{align}
as well as
\begin{align}
A^{-1}\pounds_{_{\xi}}A  &  =l_{_{0}}\sqrt{b}\nonumber\\
\pounds_{_{\xi}}\alpha_{_{0}}  &  =\frac{1}{\sin\theta}\left(  -a_{_{0}}%
\sin\phi+b_{_{0}}\cos\phi\right)  , \label{C5}%
\end{align}
where $a_{_{0}},b_{_{0}},c_{_{0}},l_{_{0}},l_{_{1}},h_{_{0}}$ are six
arbitrary integration constants. Thus,
\begin{align}
\xi &  =\left(  -\frac{l_{_{0}}}{r}+h_{_{0}}\right)  \frac{\partial}{\partial
t}+\frac{1}{2}r^{2}\left(  2l_{_{0}}bt+l_{_{1}}\right)  \frac{\partial
}{\partial r}+\left(  a_{_{0}}\cos\phi+b_{_{0}}\sin\theta\right)
\frac{\partial}{\partial\theta}\nonumber\\
&  +\left(  c_{_{0}}+\cot\theta\left(  -a_{_{0}}\sin\phi+b_{_{0}}\cos
\phi\right)  \right)  \frac{\partial}{\partial\phi}. \label{C6}%
\end{align}

Systematically putting all but one constant equal to zero we obtain the two
CKVs
\begin{align}
\xi_{_{\left(  1\right)  }}  &  =-\frac{1}{r}\frac{\partial}{\partial t}%
+r^{2}bt\frac{\partial}{\partial r}\hspace{2cm}\left(  l_{_{0}}=1,\hspace
{1cm}\varphi=2rbt\right) \nonumber\\
\xi_{_{\left(  2\right)  }}  &  =\frac{r^{2}}{2}\frac{\partial}{\partial
r}\hspace{3.8cm}\left(  l_{_{1}}=1,\hspace{1cm}\varphi=r\right)  \label{A54}%
\end{align}
as well as the four KVs
\begin{align}
\xi_{_{\left(  3\right)  }}  &  =\frac{\partial}{\partial t}\hspace
{5cm}\left(  h_{_{0}}=1\right) \nonumber\\
\xi_{_{\left(  4\right)  }}  &  =\frac{\partial}{\partial\phi}\hspace
{5cm}\left(  c_{_{0}}=1\right) \nonumber\\
\xi_{_{\left(  5\right)  }}  &  =\cos\phi\frac{\partial}{\partial\theta}%
-\cot\theta\sin\phi\frac{\partial}{\partial\phi}\hspace{1.5cm}\left(  a_{_{0}%
}=1\right) \nonumber\\
\xi_{_{\left(  6\right)  }}  &  =\sin\phi\frac{\partial}{\partial\theta}%
+\cot\theta\cos\phi\frac{\partial}{\partial\phi},\hspace{1.5cm}\left(
b_{_{0}}=1\right)  \label{A55}%
\end{align}
two of which are obvious since the metric is independent of $t$ and $\phi$.
There is no HV since $\varphi$ cannot be a nonzero constant. These symmetries
have also been obtained by Koutras\cite{Koutras} and by O'Connor and
Prince\cite{Prince2}.

In retrospect, we see that the phase part of the gauge cannot be fixed
intrinsically; therefore, there is partial gauge isotropy. For, if it could
be, this would imply that there is a common preferred phase for all four KV.
That there is none follows from the second of Eqs.(\ref{C5}). We have,
\begin{equation}
\pounds_{_{\xi_{_{_{3}}}}}\alpha_{_{0}}=0=\pounds_{_{\xi_{_{4}}}}\alpha_{_{0}%
}, \label{A56}%
\end{equation}
implying that $\alpha_{_{0}}=\alpha_{_{0}}\left(  r,\theta\right)  $. But, by
the same equation, to get a preferred phase for $\xi_{_{5}}$ as well, we need
\begin{equation}
\pounds_{_{\xi_{_{_{5}}}}}\alpha_{_{0}}=-\frac{\sin\phi}{\sin\theta},\text{
i.e. }\cos\phi\frac{\partial\alpha_{_{0}}\left(  r,\theta\right)  }%
{\partial\theta}=-\frac{\sin\phi}{\sin\theta}. \label{A57}%
\end{equation}
This cannot be satisfied.

On the other hand, there is a boost $A$ so that the resultant boost part of
the gauge is preferred for all KV. In fact, $A=1$ will do since the first of
Eqs.(\ref{C5}) reduces to $\pounds_{_{\xi}}A=0$ for the four KV (and,
therefore, there is no boost isotropy). That there is no common boost $A$
which will satisfy the first of Eq.(\ref{C5}) for all six (C)KV is readily verified.

This example illustrates how our approach deals with gauge isotropy. We do not
need to know about possible gauge isotropy in advance; whatever the reason why
a tetrad's gauge has not been fixed completely (or even if we are unsure
whether it is fixed or not) we simply introduce the gauge parameters into the
commutators, and the result of our calculation tells us whether our original
gauge was completely preferred or not; and if the latter, it tells us what
transformation takes the NP tetrad to a preferred gauge and what isotropy
exists, if any.\medskip

\textbf{Example 4\medskip\medskip}

The Kerr-Newman metric\cite{KN} is of Petrov type D with an electromagnetic
energy momentum tensor, and one standard version\cite{Bose} is\newpage%
\begin{align}
ds^{2}  &  =\left(  1-\frac{2mr-e^{2}}{\Sigma}\right)  du^{2}%
+2dudr+\frac{2a\sin^{2}\theta}{\Sigma}\left(  2mr-e^{2}\right)  dud\phi
-2a\sin^{2}\theta drd\phi\nonumber\\
&  -\Sigma d\theta^{2}-\sin^{2}\theta\left(  r^{2}+a^{2}+\frac{a^{2}\sin
^{2}\theta}{\Sigma}\left(  2mr-e^{2}\right)  \right)  d\phi^{2}, \label{H1}%
\end{align}
where $\Sigma=r^{2}+a^{2}\cos^{2}$ $\theta$ and $\Delta=r^{2}+a^{2}+e^{2}%
-2mr$. In the tetrad (with coordinates $\left(  u,r,\theta,\phi\right)  $)
\begin{align}
l^{^{\alpha}}  &  =\left(  0,1,0,0\right) \nonumber\\
n^{^{\alpha}}  &  =\left(  \frac{r^{2}+a^{2}}{\Sigma},-\frac{\Delta}{2\Sigma
},0,\frac{a}{\Sigma}\right) \nonumber\\
m^{^{\alpha}}  &  =\frac{1}{\sqrt{2}\left(  r+ia\cos\theta\right)  }\left(
ia\sin\theta,0,1,\frac{i}{\sin\theta}\right)  \label{H2a}%
\end{align}
the only nonzero tetrad components of the Riemann tensor are
\begin{align}
\Psi_{_{2}}  &  =-\frac{m\left(  r+ia\cos\theta\right)  -e^{2}}{\left(
r-ia\cos\theta\right)  ^{3}\left(  r+ia\cos\theta\right)  }\nonumber\\
\Phi_{_{11}}  &  =\frac{e^{2}}{2\Sigma^{2}}. \label{H3}%
\end{align}

The tetrad is canonical, but it is not immediately clear how and to what
extent the gauge has been fixed. We apply Theorem 21 to investigate all (C)KVs.

First, we consider how to obtain simple Lie recurrent relations for the
coordinates $r$, $\theta$, using the second half of the theorem. Since we wish
to consider all (C)KV possibilities together, using (a), we need to focus on
NP scalars which are gauge invariant and properly conformally weighted:
$\Psi_{_{2}}$, $(\rho-\overline{\rho})(\mu-\overline{\mu})$, $(\tau
+\overline{\pi})(\overline{\tau}+\pi)$. The simplest approach is to consider a
combination which is gauge invariant, and also of conformal weight zero; e.g.
$\pounds_{_{\xi}}((\rho-\overline{\rho})(\mu-\overline{\mu})(\tau
+\overline{\pi})(\overline{\tau}+\pi)/\Psi_{_{2}})=0$. When we work this
expression out explicitly we obtain, from its real and imaginary parts, two
simultaneous homogenous equations in $\pounds_{_{\xi}}r$ and $\pounds_{_{\xi}%
}\theta$; since the determinant is non-zero, we conclude that $\pounds_{_{\xi
}}r=0=\pounds_{_{\xi}}\theta$. Substituting this result into $\pounds_{_{\xi}%
}\Psi_{_{2}}=-\varphi\Psi_{_{2}}$ gives $\varphi=0$; this means that the
Kerr-Newman metric does not permit an HV or CKVs. This example illustrates how
we can obtain results on the non-existence of CKVs very efficiently, simply by
testing the consistency of the NP scalars under conformal transformations.

To investigate the KVs only we could apply Theorem 21b to $\Phi_{_{11}}$ and
$\rho$ (or to $\Psi_{_{2}}$) and again obtain easily that $\pounds_{_{\xi}%
}r=0=\pounds_{_{\xi}}\theta$. It then remains to solve the NP-Lie commutator
equations, with arbitrary gauge factor $A\exp\left(  \frac{i}{2}\alpha_{_{0}%
}\right)  $, applied to the remaining two coordinates $u$ and $\phi$. It is
easily seen that one possible solution for the gauge parameters is $A=1$,
$\alpha_{_{0}}=0$, showing that the original gauge is preferred relative to
all KVs. The solution for the Lie derivatives of $u$ and $\phi$ is
$\pounds_{_{\xi}}u=c_{_{1}\text{, }}\pounds_{_{\xi}}\phi=c$, thereby yielding
the two familiar KVs, $\frac{\partial}{\partial t}$ and $\frac{\partial
}{\partial\varphi}$.

\subsection{\textbf{PETROV TYPE N} with a canonical tetrad}

In Petrov type N a canonical tetrad is defined by $\Psi_{_{4}}=1,$ all other
$\Psi$ zero.

Such a tetrad has one null direction preferred, namely the principal null
direction, but the second null direction is not fixed. The gauge is fixed with
respect to the Weyl tensor, but not in a proper conformally weighted manner.
In practice, sometimes the null rotation freedom is fixed either by the spin
coefficients or by the Riemann tensor; but there are some subclasses of
metrics for which it is either difficult to determine a second null vector
(because of complicated calculations), or even impossible to fix with respect
to the Riemann tensor and its derivatives (because of null rotation isotropy).

If we can see that the null rotation freedom has been fixed intrinsically and
in a proper conformally weighted manner, then the analysis is carried out as
in Petrov type III with the introduction of only one arbitrary parameter ---
for a boost. But in all other cases, besides an arbitrary boost parameter $A$,
it is necessary to introduce the arbitrary complex null rotation parameter $z$
explicitly into the commutators in order to obtain a transformation to a
preferred tetrad.

In this case we can only use Lemma 14 for those scalars which are identical in
both the canonical and preferred tetrad, i.e., scalars which are invariant
under null rotations, which have proper conformal weight and which, in
addition, are boost-invariant.

\begin{theorem}
If a given metric is of Petrov N and if an NP tetrad for this metric is in the
above canonical form then, after a null rotation with parameter $z$ and a
boost with parameter $A$ to a new tetrad $\widehat{l},\widehat{n},\widehat
{m},\widehat{\overline{m}}$, all (C)KVs can be found by applying $[\widehat
{D},\pounds]=\frac{1}{2}\varphi\widehat{D}$ and its companion equations to the
four coordinates $x^{i}$, and solving for $\pounds_{_{\xi}}$ $x^{i}$, the
unknown gauge parameter $A$, and the unknown null rotation parameter $z$ and
the conformal function $\varphi$. Furthermore, (a) all null rotation
invariant, boost-invariant and proper conformally weighted combinations of the
NP scalars are Lie recurrent relative to each (C)KV, (b) all null rotation
invariant and boost-invariant weighted combinations of the NP scalars are Lie
recurrent relative to an\textbf{\ }HV and any KV.
\end{theorem}

\begin{corollary}
The \emph{KVs only} can be found by putting $A=1,\varphi=0$, and solving the
commutator equations for $\pounds_{_{\xi}}x^{i}$ and the unknown null rotation
parameter $z$. Furthermore, all null rotation invariant combinations of the NP
scalars are Lie derived relative to each KV.
\end{corollary}

\subsection{{\protect\Large \smallskip}PETROV TYPE O}

For Petrov type 0 spacetimes another approach has to be used --- usually using
the Ricci tensor to fix part of the tetrad --- and although we can still
handle those cases with the general principles described in this paper, we
will postpone such considerations to a subsequent paper.\newpage

\subsection{\smallskip ALL TYPES}

We emphasise that in practical calculations involving NP tetrads, using
canonical tetrads is not the only way in which NP tetrads can be chosen. For
instance, it is useful in some circumstances for the NP tetrad to be partly
fixed with respect to the Weyl tensor and partly with respect to the Ricci
tensor. Also, in investigations concerning the IC of metrics, canonical
tetrads are obtained which have more complicated relationships (involving
derivatives of the Riemann tensor) with the Weyl and Ricci tensors. Therefore,
in Petrov types III, D, N where we have introduced arbitrary parameters, an
alternative procedure may be to first of all find an explicit suitable gauge
transformation or null rotation, in order to avoid using some of the arbitrary
parameters --- in a manner analogous to the one we have used our discussion of
Petrov type II. But this will require a deeper analysis of the tetrad, an
analysis which involves determining how the tensor is fixed with respect to
derivatives of the Riemann tensor, as in the IC. However, we point out that
although we should always be able to eliminate the gauge parameter for Petrov
type III, we cannot always eliminate the parameters in Petrov types D and N
because of isotropy.

Our general approach outlined in the previous subsections is designed to
include as well tetrads other than the ones discussed above, and we shall
investigate such situations in a subsequent paper. However, using canonical
tetrads is a familiar approach, and in this paper we wished to illustrate the
efficiency of our method with familiar examples.

\section{Conclusion}

The main thrust of this paper was to present a very efficient, alternative,
approach to the problem of finding (conformal) Killing vectors, i.e. KVs, HV,
or CKVs. Although our examples have dealt with finding such vectors for a
given metric, our technique works equally well when solving Einstein's
equations for a spacetime with such symmetries.

This technique consists of the replacement of the usual (conformal) Killing
equations with sixteen equations involving commutators of the Lie derivative
with the four NP differential operators applied to the four coordinates. These
operators must correspond to a tetrad which is preferred relative to the (C)KV
to be found. This notion was defined for null directions in an earlier
paper\cite{EL2} and extended to gauge in this one. In practice, we usually
need a tetrad which is preferred relative to all KVs and/or HV and/or CKVs of
the metric. To cover all possible (C)KVs it is best to find null directions
which are intrinsically defined by the Weyl tensor alone (in a proper
conformally weighted manner). As for gauge, if we want to find only all KVs it
suffices to choose one which has been defined intrinsically. To get CKVs as
well, we need to look for a gauge which has been defined not only
intrinsically but in a proper conformal manner as well.

In such a preferred gauge the calculations for the (C)KVs become relatively
simple, especially if as many coordinates as possible are chosen intrinsically
as well. In case of isotropy, or if ignorant of a tetrad which is preferred
relative to all the (C)KVs we want to find, we must put an arbitrary null
rotation parameter and/or an arbitrary gauge factor into the tetrad used. Then
the calculations become a bit longer. The NP-Lie commutator equations must now
be solved for the Lie derivative of such a factor as well as for the Lie
derivatives of the coordinates. Clearly, in the interest of simplicity of
calculations, we should choose as much of the tetrad to be preferred relative
to all vectors of interest, so as to minimize the use of null rotation and/or
boost-rotation parameters. This can often be done in the manner described and
illustrated in section 4. The choice depends on whether we're looking only for
KVs, for an HV and KVs, or for CKVs\ as well. Further simplification is
achieved by choosing as many coordinates as possible in an intrinsic fashion.
All the KVs, any HV, and all the CKVs are then readily obtained.

Our results for (C)KVs apply to proper conformal Killing vectors, homothetic
vectors and Killing vectors. It might be thought superfluous for us to give
specific results valid only for KVs or an HV. However, as noted in the
introduction, there are some general theorems which, in certain circumstances,
rule out the existence for CKVs. In such circumstances it is more efficient to
use the results which are valid only for KVs and/or an HV.

The examples which we have used to illustrate our method are comparatively
simple, being metrics quoted with tetrads in canonical form. However, our
method is applicable, and efficient, for other types of tetrad. In particular,
our method can be used to extend the IC program by determining explicitly the
form of those KVs and HVs whose existence can be concluded from the existing
IC program. In addition, our results will enable the IC program to deal with
HVs and CKVs. These applications will be published in a subsequent paper.

\bigskip

\begin{description}
\item {\Large ACKNOWLEDGEMENTS}
\end{description}

One author (BE) gratefully acknowledges the financial support by the Natural
Science Research Council of Sweden, the travel support from the Magnusson
Fund, Swedish Royal Academy, and the hospitality of the Department of
Mathematical Sciences, University of Alberta, Canada. The other author (GL)
would like to thank the Natural Sciences and Engineering Research Council of
Canada for its ongoing financial support.


\begin{thebibliography}{9}                                                                                                %

\bibitem {EL1}Edgar S.B. and Ludwig G. (2000). \textit{Gen. Rel. Grav.
}\textbf{32}, 637-671.

\bibitem {EL2}Ludwig G. and Edgar S. B. (2000). \textit{Class. Quantum Grav}.
\textbf{17}, 1683-1705.

\bibitem {GHP}Geroch R., Held A., and Penrose R. (1973). \textit{J. Math.
Phys.} \textbf{14}, 874-881.

\bibitem {NP}Newman E.T. and Penrose R. (1962). \textit{J. Math. Phys.}
\textbf{3}, 566-578.

\bibitem {Karl}Karlhede A. (1980). \textit{Gen. Rel. Grav. \textbf{12}, }693-707.

\bibitem {MK}Karlhede A. and MacCallum M.A.H.(1982). \textit{Gen. Rel. Grav.
\textbf{14},} 673-682.

\bibitem {KS}Koutras A. and Skea J.E.F. (1998). \textit{Computer Physics
Communications} \textbf{115}, 350-362.

\bibitem {Hall1}Hall G.S. (2000). \textit{Gen. Rel. Grav. }\textbf{32}, 933-941.

\bibitem {Hall2}Hall G.S. (2000). \textit{Class. Quantum Grav}. \textbf{17}, 3073-3076.

\bibitem {Prince}Jerie M., O'Connor, J.E.R. and Prince G.E. (1999).
\textit{Class. Quantum Grav}. \textbf{16}, 2885-2887.

\bibitem {KL}Kolassis Ch. and Ludwig G. (1993). \textit{Gen. Rel. Grav.}
\textbf{25}, 625-651.

\bibitem {L1}Ludwig G. (1988). \textit{Int. J. Theor. Phys. }\textbf{27}, 315-333.

\bibitem {MV1}Machado Ramos M.P. and Vickers J.A. (1996). \textit{Class.
Quantum Grav}. \textbf{13}, 1579-1587.

\bibitem {BS}Bell P. and Szekeres P. (1972). \textit{Int. J. Theor. Phys.}
\textbf{6}, 111-121.

\bibitem {Hall}Hall G.S. and Keane A.J. (2000). \textit{Class. Quantum Grav}.
\textbf{17}, 1571-1573.

\bibitem {Steele}Steele J.D. (1991). \textit{Gen. Rel. Grav. }\textbf{23}, 811-825.

\bibitem {Kimura}Kimura M. (1976). \textit{Tensor} \textbf{30}, 27-43.

\bibitem {Koutras}Koutras A. (1992). \textit{Class. Quantum Grav}. \textbf{9}, 1573-1580.

\bibitem {Prince2}O'Connor, J.E.R. and Prince G.E. (1998). \textit{Gen. Rel.
Grav. }\textbf{30}, 69-82.

\bibitem {KN}Newman E.T. et al. (1965). \textit{J. Math. Phys.} \textbf{6}, 918-919.

\bibitem {Bose}Bose S.K. (1975). \textit{J. Math. Phys.} \textbf{16}, 772-775.
\end{thebibliography}
\end{document}